\documentclass[aps]{revtex4-2}
\usepackage{amsmath}
\usepackage{amssymb}
\usepackage{graphicx}
\usepackage{geometry}
\usepackage{booktabs} 
\usepackage{hyperref} 
\usepackage[utf8]{inputenc}
\usepackage{float}

\begin{document}
\title{Non-Thermal Leptogenesis in the BLSM with Inverse Seesaw Mechanism}
\author{David Delepine}
\email{delepine@ugto.mx}
\affiliation{{\fontsize{10}{10}\selectfont{Division de Ciencias e Ingenier\'ias,  Universidad de Guanajuato, C.P. 37150, Le\'on, Guanajuato, M\'exico.}}}
\author{Shaaban Khalil}
\email{skhalil@zewailcity.edu.eg}
\affiliation{{\fontsize{10}{10}\selectfont{Centre for Theoretical Physics, Zewail City of Science and Technology, 6th October City, 12588, Giza, Egypt.}}}

\begin{abstract}
We investigate the viability of non-thermal leptogenesis in the gauged
$U(1)_{B-L}$ extension of the Standard Model (BLSM) with an inverse seesaw (ISS)
mechanism for neutrino mass generation. In this framework, right-handed
neutrinos typically have $\mathcal{O}(1)$ Yukawa couplings, which induce
strong washout effects and render conventional thermal leptogenesis
ineffective. We demonstrate that a successful baryogenesis scenario can
nevertheless be realized through non-thermal leptogenesis, where
right-handed neutrinos are produced from the decay of the heavy
$B\!-\!L$ Higgs boson $\chi$. We explicitly analyze the interplay between
the dilution factor $T_R/M_\chi$ and the washout parameter characteristic
of the ISS, highlighting the tension between suppressing washout effects
and maintaining sufficient reheating. We show that a viable lepton
asymmetry can be generated provided the scalar mass spectrum is
appropriately tuned --- a condition that requires a fine-tuning of order
one part in $10^5$ against one-loop radiative corrections ---
allowing for a reduced reheating temperature while keeping washout under
control. The resulting lepton asymmetry is
efficiently converted into the observed baryon asymmetry of the Universe
via sphaleron processes. Our results establish that the inverse-seesaw
$B\!-\!L$ model remains a predictive and robust framework for non-thermal
leptogenesis and baryogenesis.
\end{abstract}

\maketitle
\section{Introduction}

The origin of the baryon asymmetry of the Universe (BAU) remains one of the most fundamental open problems in particle physics and cosmology. Observations of the cosmic microwave background and primordial nucleosynthesis indicate a small but nonzero baryon-to-photon ratio \cite{Planck:2018vyg,Workman:2022ynf},
\begin{equation}
\eta_B^{\rm obs} \simeq 6 \times 10^{-10},
\end{equation}
which cannot be explained within the Standard Model (SM). Any successful mechanism for baryogenesis must satisfy the Sakharov conditions \cite{Sakharov:1967dj}: baryon number violation, $C$ and $CP$ violation, and a departure from thermal equilibrium. Among the various proposals, leptogenesis provides an attractive and economical framework in which a lepton asymmetry is first generated and later converted into a baryon asymmetry through electroweak sphaleron processes.

In its conventional realization, thermal leptogenesis relies on the out-of-equilibrium decay of heavy Majorana neutrinos that are thermally produced in the early Universe. Successful baryogenesis in this framework typically requires heavy-neutrino masses well above the TeV scale, $M_N \gtrsim 10^9$~GeV \cite{Davidson:2002qv}, together with sufficiently small washout effects \cite{Fukugita:1986hr,Giudice:2003jh,Davidson:2008bu}. While theoretically appealing, such high scales are difficult to probe experimentally and are in tension with low-scale realizations of neutrino mass generation. This has motivated extensive efforts to construct leptogenesis scenarios that operate at or near the TeV scale, where a closer connection to neutrino physics and collider phenomenology can be established \cite{Boucenna:2014uma,Pilaftsis:2005rv,Akhmedov:1998qx,Asaka:2005pn,DAmbrosio:2003wy,Deppisch:2010fr}.

Non-thermal leptogenesis provides a well-studied alternative to the thermal paradigm \cite{Lazarides:1991wu}. In this class of scenarios, the heavy right-handed neutrinos are produced out of equilibrium from the decay of a heavier field, most notably the inflaton \cite{Asaka:1999yd}. In inflaton-induced non-thermal leptogenesis, the inflaton couples directly to right-handed neutrinos and decays into them after inflation, generating a lepton asymmetry through their subsequent decays. This mechanism naturally avoids strong thermal washout effects and allows successful baryogenesis even for reheating temperatures well below the heavy-neutrino mass, $T_R \ll M_N$. Such scenarios have been widely investigated in a variety of contexts, including minimal seesaw models, supersymmetric frameworks, and resonant leptogenesis, and have been shown to reproduce the observed BAU under well-defined assumptions on the inflaton sector, its couplings, and the reheating dynamics \cite{Asaka:1999yd,Giudice:1999fb, Kumekawa:1994gx,Hahn-Woernle:2008tsk}.

While inflaton-decay scenarios provide a compelling proof of principle for non-thermal leptogenesis, their assumptions about the inflationary sector and its direct couplings to the neutrino sector are practically not constrained from a particle-physics perspective. So, the exploration of alternative non-thermal mechanisms in which the source of right-handed neutrinos arises from well-defined and testable degrees of freedom associated with extensions of the SM gauge structure is still an interesting questions. 

A particularly well-motivated framework in this regard is the gauged $U(1)_{B-L}$ extension of the Standard Model. In this theory, the existence of three right-handed neutrinos is required by anomaly cancellation, and light neutrino masses can be generated naturally via the inverse seesaw (ISS) mechanism \cite{Mohapatra:1986bd,Khalil:2010iu}. The ISS allows the heavy neutrinos to reside at the TeV scale while accommodating realistic light neutrino masses with moderately large Yukawa couplings. As a result, the $B-L$ inverse seesaw model is highly predictive and experimentally testable \cite{Khalil:2006yi,Khalil:2007dr,Abdallah:2011ew,Elsayed:2011de}. However, these same features lead to a severe cosmological challenge: the large Yukawa couplings induce extremely strong washout effects, rendering conventional thermal leptogenesis catastrophically ineffective, even when resonant $CP$ enhancement is taken into account.

This observation strongly suggests that the origin of the lepton asymmetry in the $B-L$ inverse seesaw framework must be non-thermal. In this work, we show that a baryogenesis scenario can be realized through non-thermal leptogenesis driven by the decay of the scalar field responsible for $B-L$ symmetry breaking. In this setup, the heavy (pseudo-Dirac) neutrinos are produced out of equilibrium from the decay of the $B-L$ Higgs boson, while the reheating temperature remains below the heavy-neutrino mass, inducing  an exponential suppression of inverse decays and washout processes.

We show that the interplay of three key ingredients allows the observed baryon asymmetry to be reproduced: (i) non-thermal production of heavy neutrinos from $B-L$ Higgs decay, (ii) a naturally suppressed reheating temperature achieved when the scalar mass lies close to the kinematic threshold for neutrino production, and (iii) resonant enhancement of the $CP$ asymmetry enabled by the small lepton-number-violating parameter intrinsic to the inverse seesaw.

The paper is organized as follows. In Section~II, we review the $U(1)_{B-L}$ inverse seesaw model and show the failure of thermal leptogenesis due to extreme washout. Section~III presents the non-thermal leptogenesis mechanism arising from $B-L$ Higgs decay and discusses the parametric conditions required for suppressing washout while maintaining a sufficient heavy-neutrino yield. Sections~IV and~V analyze the washout regimes and resonant $CP$ violation in detail. Our numerical results are presented in Section~VI, and we summarize our conclusions in Section~VII.

\section{BLSM-ISS and the Failure of Thermal Leptogenesis}

The gauged $U(1)_{B-L}$ extension of the SM provides a minimal and
well-motivated framework in which three right-handed neutrinos (RHN) arise naturally
from anomaly cancellation. This setup allows also for a low-scale realization of the
inverse seesaw (ISS) mechanism, where light neutrino masses are generated without
requiring ultra-heavy Majorana states.
The leptonic sector relevant for neutrino mass generation is described by the
Lagrangian~\cite{Khalil:2010iu}
\begin{align}
\mathcal{L}_{B-L} &\supset 
i \bar{\ell}_L \gamma^\mu D_\mu \ell_L
+ i \bar{e}_R \gamma^\mu D_\mu e_R
+ i \bar{N}_R \gamma^\mu D_\mu N_R
+ i \bar{S}_{1,2} \gamma^\mu D_\mu S_{1,2}
+ (D_\mu \phi)^\dagger D^\mu \phi
+ (D_\mu \chi)^\dagger D^\mu \chi \nonumber\\
&\quad -
V(\phi, \chi) - \Big( 
\lambda_\nu \bar{\ell}_L \tilde{\phi} N_R 
+ \lambda_N \bar{N}_R^c \chi S_2
+ \text{h.c.} \Big)
- \frac{\alpha}{M^3}\bar{S}^c_{1} {\chi^\dag}^{4} S_{1}-\frac{\beta}{M^3}\bar{S}^c_{2} {\chi}^{4} S_{2},
\end{align}
where $\phi$ is the SM Higgs doublet and $\chi$ is the scalar field
responsible for $B\!-\!L$ symmetry breaking. The parameters $\alpha$ and $\beta$ are free coupling constants. 
Furthermore, in order to forbid a potentially large mass term $\mu \, S_1 S_2$ in the above Lagrangian, we assume that the particles, $N_R$, $\chi$, and $S_2$ are even under matter parity, while $S_1$ is odd. Finally, $V(\phi,\chi)$ is the most general Higgs potential invariant under these symmetries, which can be found in Ref.~\cite{Khalil:2006yi}.
Radiative $B\!-\!L$ breaking in supersymmetric realizations typically yields
$v'=\langle\chi\rangle\sim\mathcal{O}(\mathrm{TeV})$~\cite{Khalil:2007dr}.

After symmetry breaking, the inverse-seesaw neutrino mass matrix in the
$(N_R,S_2)$ basis reads
\begin{equation}
\mathcal{M}_\nu=
\begin{pmatrix}
0 & M_N \\
M_N^T & \mu_S
\end{pmatrix},
\qquad
M_N=\frac{\lambda_N v'}{\sqrt{2}},
\end{equation}
where $\mu_S = \beta v'^4/M^3 \sim 10^{-9}\,\text{GeV}$. Consequently, one obtains quasi-Dirac heavy neutrino states with masses
\begin{equation}
m_{\nu_{H,H'}} \simeq \pm M_N + \frac{\mu_S}{2},
\qquad
\nu_{H,H'} \simeq \frac{1}{\sqrt{2}}\left(\mp N_R + S_2\right).
\end{equation}

The light neutrino mass matrix generated via the inverse seesaw takes the form
\begin{equation}
m_\nu \simeq \frac{(y_\nu v_{\rm SM})^2}{M_N^2}\,\mu_S ,
\end{equation}
where $v_{\rm SM}\simeq174$~GeV. Reproducing the atmospheric neutrino mass scale
$\sqrt{\Delta m^2_{\rm atm}}\simeq0.05$~eV implies
\begin{equation}
y_\nu \simeq \frac{M_N}{v_{\rm SM}}
\sqrt{\frac{0.05~\mathrm{eV}}{\mu_S}} ,
\end{equation}
showing that for TeV-scale $M_N$ and small lepton-number-violating parameter
$\mu_S$, the Dirac Yukawa couplings are naturally of
$\mathcal{O}(10^{-1}\!-\!1)$.

\subsection*{CP Asymmetry and Resonant Enhancement}

The CP asymmetry generated in the decay of the lightest heavy neutrino
$\nu_{H_1}$ arises from the interference between tree-level and one-loop
diagrams~\cite{Fukugita:1986hr,Buchmuller:1997yu,Covi:1996wh}:
\begin{equation}
\epsilon_{CP} \simeq 
\frac{1}{4\pi}\,
\frac{\sum_{k=2,3}
\mathrm{Im}\!\left[(\lambda_\nu\lambda_\nu^\dagger)_{1k}^2\right]
F(x_k)}
{(\lambda_\nu\lambda_\nu^\dagger)_{11}},
\qquad
x_k=\frac{m_{\nu_{H_k}}^2}{m_{\nu_{H_1}}^2},
\end{equation}
with
\begin{equation}
F(x)=\sqrt{x}\!
\left[
1+\frac{1}{1-x}-(1+x)\ln\frac{1+x}{x}
\right].
\end{equation}
In pseudo-Dirac systems such as the Inverse Seesaw, the interference term driving the CP asymmetry can be highly suppressed if the flavor structure is aligned. However, by introducing complex phases in the orthogonal matrix of the Casas-Ibarra parameterization, these catastrophic cancellations are  avoided. We provide a concrete flavor realization and the explicit calculation of our benchmark asymmetry ($\epsilon_{CP} \simeq 0.48$) in Appendix \ref{sec:appendix_A}, ensuring full compatibility with low-energy neutrino oscillation data.

In the ISS, the heavy neutrinos form pseudo-Dirac pairs with a small mass
splitting governed by $\mu_S$,
\begin{equation}
\Delta M \simeq \mu_S .
\end{equation}
Resonant enhancement of CP violation occurs when
\begin{equation}
\Delta M \sim \frac{\Gamma}{2},
\end{equation}
in which case the CP asymmetry can reach its theoretical maximum,
$\epsilon_{\rm CP}\lesssim 0.5$ \cite{Pilaftsis:1997jf,Pilaftsis:2003gt, Pilaftsis:2005rv,Covi:1996wh,Flanz:1996fb,Anisimov:2005hr}. However, due to the pseudo-Dirac structure,
the asymmetries generated by the two components partially cancel, making the
net enhancement highly model-dependent.

\subsection*{Strong Washout and Breakdown of Thermal Leptogenesis}

The efficiency of thermal leptogenesis is governed by the decay parameter
\begin{equation}
K=\frac{\Gamma_{D_1}}{H}\Big|_{T=M_N},
\qquad
Y_B \simeq -1.4\times10^{-3}\,\kappa(K)\,\epsilon ,
\end{equation}
where $\kappa(K)$ encodes washout effects~\cite{Buchmuller:2004nz}. The Hubble
rate in standard cosmology is
\begin{equation}
H \simeq 1.66\,\sqrt{g_\ast}\,\frac{T^2}{M_{\rm Pl}},
\qquad g_\ast\simeq\mathcal{O}(10^2).
\end{equation}

For TeV-scale heavy neutrinos with $y_\nu=\mathcal{O}(1)$, one finds
\begin{equation}
H \sim 10^{-12}\ \mathrm{GeV},
\qquad
K \sim 10^{12}\!-\!10^{14},
\end{equation}
placing the system deep in the strong-washout regime. In this limit:
\begin{itemize}
\item Inverse decays ($LH \rightarrow N$) remain fully active in the thermal bath.
\item The efficiency factor is exponentially suppressed,
$\kappa(K)\sim10^{-15}$.
\item Even with maximal resonant CP asymmetry,
$\epsilon_{\rm CP}\sim0.5$, the resulting baryon asymmetry is
$\eta_B\sim10^{-16}$, far below the observed value
$\eta_B^{\rm obs}\simeq6\times10^{-10}$.
\end{itemize}

We therefore conclude that \emph{thermal leptogenesis is catastrophically
ineffective in the $B\!-\!L$ inverse seesaw model}. This failure is driven by
the unavoidable combination of large Yukawa couplings and extreme washout,
independently of possible resonant CP enhancement.

This observation  motivates non-thermal leptogenesis scenarios, in which
right-handed neutrinos are produced out of equilibrium \cite{Racker:2008sg,Blanchet:2009bu,Blanchet:2009iz,Abdallah:2012nm}, through the
decay of the heavy $B\!-\!L$ Higgs scalar.


\section{Non-thermal leptogenesis from $B\!-\!L$ Higgs decay}

As shown in the previous section, \emph{thermal} leptogenesis is catastrophically
ineffective in the inverse-seesaw (ISS) $B\!-\!L$ framework: the Yukawa couplings
required to reproduce $m_\nu$ at the TeV scale are typically
$y_\nu=\mathcal{O}(10^{-1}\!-\!1)$, implying an enormous washout parameter
$K\sim10^{12}\!-\!10^{14}$ and hence an exponentially suppressed efficiency.
This motivates \emph{non-thermal leptogenesis}, where the heavy (pseudo-Dirac)
neutrinos are produced out of equilibrium through the decay of the $B\!-\!L$
breaking scalar $\chi$, while the reheating temperature satisfies
$T_R<M_N$ so that inverse decays are Boltzmann suppressed.

\subsection{Basic idea and parametric requirements}

Non-thermal leptogenesis avoids (or strongly reduces) the thermal problems by
engineering the following conditions:
\begin{itemize}
\item \textbf{Low reheating, $T_R<M_N$:} the Universe reheats to a temperature
below the heavy-neutrino mass, so that $LH\to N$ inverse decays and related
washout processes are kinematically disfavored and Boltzmann suppressed.
\item \textbf{Controlled washout, $K_{\rm eff}\ll K$:} the effective washout is
reduced relative to the thermal value by a factor  as
$\exp(-M_N/T_R)$, protecting the generated asymmetry.
\item \textbf{Successful yield and asymmetry:} combining an enhanced CP
asymmetry (e.g.\ resonant $\epsilon_{\rm CP}\sim0.1$--$0.3$) with a protected
efficiency $\kappa$ and a sizable non-thermal RHN yield can reproduce the
observed baryon asymmetry $\eta_B^{\rm obs}\simeq 6\times10^{-10}$.
\end{itemize}

\subsection{Reheating temperature and non-thermal RHN yield}

We assume that after inflation the energy density is dominated by coherent
oscillations of the scalar field $\chi$ (matter-dominated era) and that when $H\simeq \Gamma_\chi$, where $\Gamma_\chi$ is the total decay width of
$\chi$, reheating is taking place. The reheating temperature is then \cite{Kolb:1990vq,Kofman:1997yn,Allahverdi:2010xz}
\begin{equation}
T_R \simeq \left(\frac{90}{8\pi^3 g_*(T_R)}\right)^{1/4}\sqrt{\Gamma_\chi M_{\rm Pl}}\,,
\label{eq:TR_general}
\end{equation}
with $g_*(T_R)\sim\mathcal{O}(10^2)$.
If a fraction of $\chi$ decays produces heavy neutrinos,
$\chi\to NN$ (here $N$ collectively denotes the heavy ISS states), the RHN yield
at reheating is approximately
\begin{equation}
Y_N^{\rm non\text{-}th}\equiv \frac{n_N}{s}
\;\simeq\;
\frac{3}{2}\,{\rm Br}(\chi\to NN)\,\frac{T_R}{M_\chi}\,,
\label{eq:YN_nonthermal}
\end{equation}
where $s$ is the entropy density and ${\rm Br}(\chi\to NN)$ is the branching ratio
into heavy neutrinos. Equation~\eqref{eq:YN_nonthermal} makes explicit the
\emph{dilution} factor $T_R/M_\chi$ discussed previously: lowering $T_R$ helps
washout, but it also reduces the injected RHN abundance.


\subsection{Threshold Kinematics and the Reheating Temperature}
\label{sec:threshold_kinematics}

In our model, the heavy pseudo-Dirac neutrinos are produced out of equilibrium via the decays of the massive $B-L$ scalar field, $\chi \to NN$. To ensure that the generated asymmetry is not subsequently erased by inverse decays, the universe must never reach temperatures comparable to the heavy neutrino mass ($T_{max} \sim T_R \ll M_N$). 

This low-temperature regime is achieved by adopting a near-threshold kinematic configuration. We assume the scalar mass $M_\chi$ is just slightly above the kinematic threshold for producing two heavy neutrinos:
\begin{equation}
    M_\chi = 2M_N + \delta, \quad \text{with} \quad \delta \ll M_N.
\end{equation}

To see how this suppresses the energy injection into the thermal bath, we must  evaluate the tree-level decay width of the scalar into the heavy neutrinos. For a scalar decaying into two fermions of mass $M_N$, the width is  governed by the phase space velocity $\beta$:
\begin{equation}
    \Gamma(\chi \to NN) = \frac{Y_N^2}{8\pi} M_\chi \beta^3, \quad \text{where} \quad \beta = \sqrt{1 - \frac{4M_N^2}{M_\chi^2}}.
\end{equation}
By inserting the threshold condition $M_\chi = 2M_N + \delta$, we can expand the phase space parameter $\beta$ in the limit $\delta \ll M_N$:
\begin{equation}
    \beta = \sqrt{1 - \frac{4M_N^2}{(2M_N + \delta)^2}} \simeq \sqrt{1 - \frac{4M_N^2}{4M_N^2(1 + \delta/M_N)}} \simeq \sqrt{\frac{\delta}{M_N}}.
\end{equation}
Substituting this back into the decay width, and approximating $M_\chi \simeq 2M_N$, the partial width is  suppressed by the small mass splitting:
\begin{equation}
    \Gamma(\chi \to NN) \simeq \frac{Y_N^2}{4\pi} M_N \left( \frac{\delta}{M_N} \right)^{3/2}.
\end{equation}

To determine the total decay width of the scalar, $\Gamma_\chi$, we must consider the full scalar sector. The physical $B-L$ scalar $\chi$ can, in principle, decay into Standard Model (SM) particles (e.g., $W^+W^-$, $ZZ$, $hh$, $f\bar{f}$) through its mixing with the SM Higgs doublet. Let this mixing angle be $\alpha$. However, precision measurements of the 125 GeV Higgs boson properties at the LHC strictly require this mixing to be highly suppressed, typically $\sin \alpha \lesssim 0.1$. Because the decay widths of $\chi$ into SM states are proportional to $\sin^2 \alpha$, these channels are heavily penalized.

Consequently, the decay channel into the heavy neutrinos, governed by the unsuppressed $U(1)_{B-L}$ coupling $Y_N$, dominates. It is therefore a  correct assumption to parameterize the branching ratio into heavy neutrinos as an $\mathcal{O}(1)$ fraction of the total $\chi$ decay rate.

Finally, the total decay width  dictates the reheating temperature $T_R$ of the universe. Approximating the onset of radiation domination as the moment when the Hubble parameter equals the scalar decay rate ($H \simeq \Gamma_\chi$), and using the standard radiation density relation $H = \sqrt{\pi^2 g_* / 90} \, T^2 / M_{Pl}$,  the reheating temperature is given by:
\begin{equation}
    T_R = \left( \frac{90}{8 \pi^3 g_*} \right)^{1/4} \sqrt{\Gamma_\chi M_{Pl}}.
\end{equation}
Because $\Gamma_\chi$ is strongly suppressed by the factor $(\delta/M_N)^{3/2}$, the resulting reheating temperature is naturally driven far below the heavy neutrino mass scale, $T_R \ll M_N$, dynamically shutting off the thermal washout.

This mechanism keeps $T_R$ strictly below $M_N$, which is precisely what
ensures an exponential reduction of washout. It is also a strong signatures of the non-thermal leptogenesis scenario. In this scenario, the scalar mass spectrum is needed to be appropriately tuned allowing for a reduced reheating temperature while keeping washout under control and keeping the advantage of the ISS B-L model. 

We note that maintaining the  condition $\delta \ll M_N$ requires parameter tuning against radiative corrections. The one-loop correction to the scalar mass from the heavy neutrinos is $\Delta M_\chi \simeq (Y_N^2 / 64\pi^2) M_N$. For our benchmark parameters yielding $T_R \approx 1$ TeV ($M_N = 5$ TeV, $Y_N \simeq 0.1$, $\delta = 10^{-6}$ GeV), the radiative shift is of order $10^{-2}$ GeV. So, a fine-tuning of $\sim 1$ part in $10^5$ between the bare scalar mass and the one-loop corrections is required. This tuning is  acceptable for exploring the  viability of the mechanism.

\subsection{Washout: from $K$ to an effective $K_{\rm eff}$}

For reference, the decay width of a heavy neutrino into a lepton and Higgs,
$N\to L H$, is
\begin{equation}
\Gamma_D \simeq \frac{1}{8\pi}(y_\nu^\dagger y_\nu)\,M_N,
\label{eq:GammaD}
\end{equation}
and the Hubble rate in radiation domination is
\begin{equation}
H(T)=1.66\,\sqrt{g_*}\,\frac{T^2}{M_{\rm Pl}}\,.
\label{eq:Hubble}
\end{equation}
In thermal leptogenesis one evaluates $K=\Gamma_D/H$ at $T=M_N$, obtaining
\begin{equation}
K \equiv \frac{\Gamma_D}{H}\Big|_{T=M_N}
\simeq
\frac{(y_\nu^\dagger y_\nu)\,M_{\rm Pl}}
{8\pi\cdot 1.66\,\sqrt{g_*}\,M_N}\,,
\label{eq:K_standard}
\end{equation}
which becomes enormous for $y_\nu=\mathcal{O}(10^{-1}\!-\!1)$ and $M_N\sim\text{TeV}$.

In the non-thermal case, however, the relevant washout operates around
$T\simeq T_R<M_N$. The inverse decays are then Boltzmann suppressed, and a useful
parametric estimate is \cite{Hahn-Woernle:2008tsk}
\begin{equation}
K_{\rm eff} \;\simeq\; K\;\exp\!\left(-\frac{M_N}{T_R}\right).
\label{eq:Keff}
\end{equation}
The reheating temperature
inherits the strong phase-space suppression from $\Gamma_\chi$,
\begin{equation}
T_R(\delta)\;\propto\;\sqrt{\Gamma_\chi}\;\propto\; Y_N\,\beta^{3/2}\,\sqrt{M_\chi M_{\rm Pl}}
\;\sim\; Y_N\left(\frac{\delta}{M_N}\right)^{3/4}\sqrt{M_N M_{\rm Pl}}\,,
\label{eq:TR_delta_scaling}
\end{equation}

\subsection{Baryon asymmetry and the dilution--washout tension}

In our scenario, the baryon asymmetry $\eta_B$ can be written as
\begin{equation}
\eta_B \;\simeq\; d\;\epsilon_{\rm CP}\;\kappa(K_{eff})\;Y_N^{\rm non\text{-}th},
\label{eq:etaB_master}
\end{equation}
where:
\begin{itemize}
\item $d$ is the sphaleron conversion factor (numerically $d\simeq 0.35$),
encoding the conversion of $B\!-\!L$ into $B$ by electroweak sphalerons,
\item $\epsilon_{\rm CP}$ is the CP asymmetry per heavy-neutrino decay (in the
ISS it can be resonantly enhanced when the pseudo-Dirac splitting is comparable
to the decay width),
\item $\kappa(K_{eff})$ is an efficiency factor controlled by washout, and is strongly
correlated with $K_{\rm eff}$,
\item $Y_N^{\rm non\text{-}th}$ is the non-thermal RHN yield of
Eq.~\eqref{eq:YN_nonthermal}.
\end{itemize}
Substituting Eq.~\eqref{eq:YN_nonthermal} into Eq.~\eqref{eq:etaB_master} gives
\begin{equation}
\eta_B \;\simeq\;
d\;\epsilon_{\rm CP}\;\kappa\;
\left(\frac{3}{2}\,{\rm Br}(\chi\to NN)\,\frac{T_R}{M_\chi}\right).
\label{eq:etaB_substituted}
\end{equation}
Increasing $T_R$ boosts production
linearly via $T_R/M_\chi$, but also increases washout through $K_{\rm eff}$
(roughly exponentially sensitive to $T_R$). Taking  $M_\chi = 2M_N + \delta$  allows $T_R$ to be small enough to suppress washout,
while keeping ${\rm Br}(\chi\to NN)$  sizable.

\subsection{Competing decay channels and the Higgs-portal constraint}

A successful non-thermal scenario also requires that $\chi$ decays mainly 
into heavy neutrinos. Since $\chi$ is a Higgs-like scalar in an extended sector,
it can couple to the SM Higgs through a portal interaction, which
induces $\chi\to hh$ decays after symmetry breaking. Parametrically one may write
\begin{equation}
\Gamma(\chi\to hh)\;\sim\;\frac{\lambda_{\chi H}^2\,v'^2}{32\pi\,M_\chi},
\label{eq:Gamma_chi_hh}
\end{equation}
so that the branching ratio into neutrinos is
\begin{equation}
{\rm Br}(\chi\to NN)=
\frac{\Gamma(\chi\to NN)}
{\Gamma(\chi\to NN)+\Gamma(\chi\to hh)+\cdots}\,.
\label{eq:BR_def}
\end{equation}
In the threshold regime, $\Gamma(\chi\to NN)$ is suppressed by $\beta^3$, hence
$\lambda_{\chi H}$ must be sufficiently small to prevent $\chi\to hh$ from
dominating and driving ${\rm Br}(\chi\to NN)\to 0$ \cite{Okada:2010wd,Dev:2014tpa}. This requirement is essential
for maintaining a large $Y_N^{\rm non\text{-}th}$ and, consequently, a viable
$\eta_B$.

\subsection{Benchmark illustration and figures}

\begin{figure}
    \centering
    \includegraphics[width=0.75\linewidth]{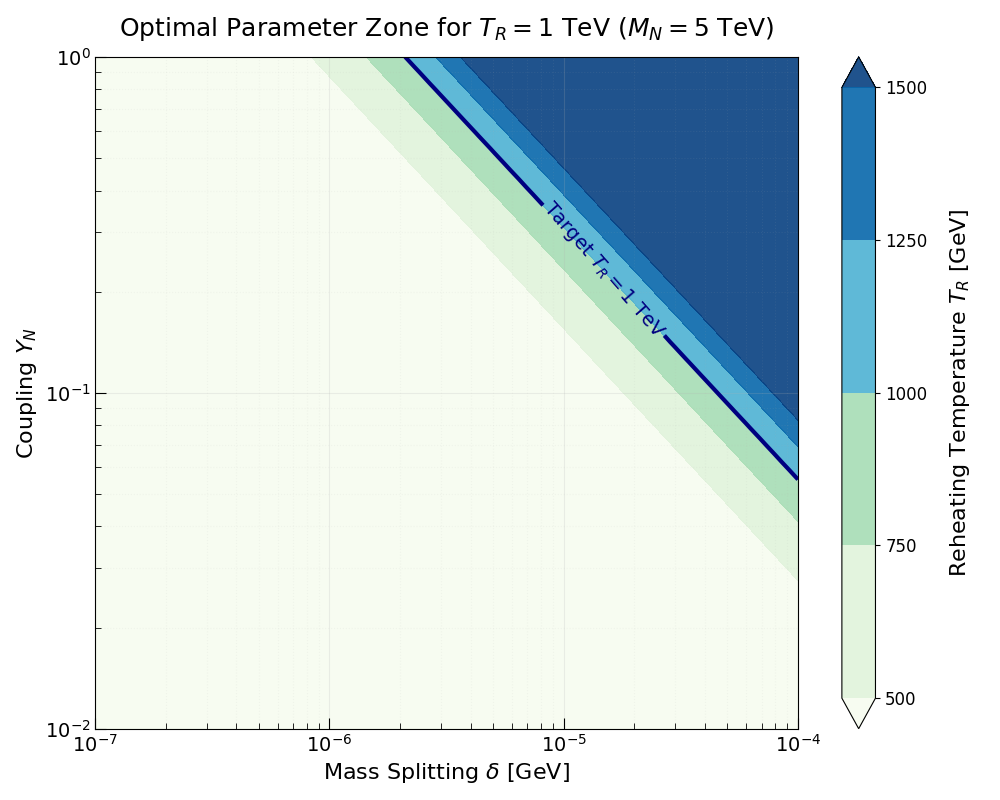}
    \caption{\textbf{Dynamical Reheating Temperature Contours.} Contour plot displaying the Reheating Temperature ($T_R$) as a function of the scalar mass splitting ($\delta$) and the scalar-neutrino coupling ($Y_N$). The scanned parameter space spans $\delta \in [10^{-7}, 10^{-4}]$ GeV and $Y_N \in [10^{-2}, 1]$. The remaining parameters are fixed to the benchmark values: $M_N = 5$ TeV, $\text{Br}(\chi \to NN) = 0.5$, and $y_\nu = 10^{-2}$.  The $T_R$ values in this plot are  computed dynamically by identifying the exact moment of radiation domination ($\rho_R = \rho_\chi$) via the full numerical integration of the coupled Boltzmann equations presented in Sec. IV. The solid navy line highlights the optimal target zone corresponding to $T_R = 1$ TeV, demonstrating that a near-threshold configuration ($\delta \sim 10^{-6}$ GeV) and a moderately large coupling ($Y_N \sim 0.1$)  yield the low reheating temperatures required to avoid thermal washout.}
\label{fig:TR_contours}
\end{figure}

\begin{figure}
    \centering
    \includegraphics[width=0.75\linewidth]{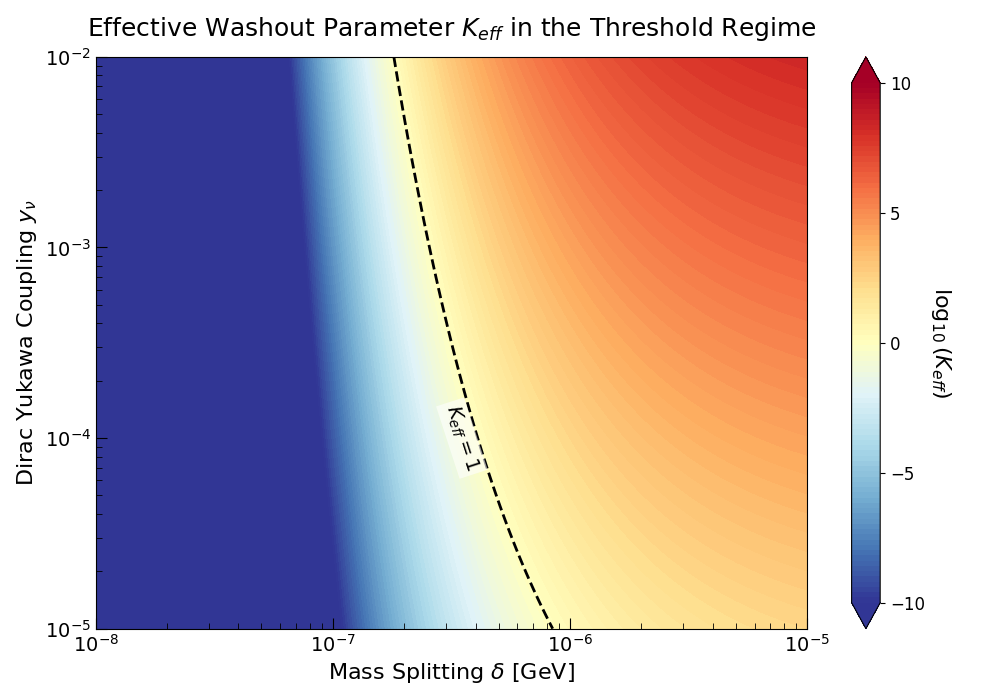}
  \caption{\textbf{Effective Washout Parameter $K_{eff}$ in the Threshold Regime.} Contour plot illustrating the logarithmic magnitude of the effective washout parameter, $\log_{10}(K_{eff})$, as a function of the scalar mass splitting $\delta$ and the Dirac Yukawa coupling $y_\nu$. The parameter space is scanned over $\delta \in [10^{-8}, 10^{-5}]$ GeV and $y_\nu \in [10^{-5}, 10^{-2}]$. The remaining parameters are fixed to $M_N = 5$ TeV and the scalar-neutrino coupling $Y_N = 1.0$. To generate this, we utilize the analytical approximation $K_{eff} \simeq K \exp(-M_N / T_R)$, where $K$ is the standard thermal washout factor evaluated at $T = M_N$, and $T_R$ is the reheating temperature analytically derived from the threshold-suppressed scalar decay width ($\Gamma_\chi \propto \beta^3 \simeq (\delta/M_N)^{3/2}$). The dashed black contour delineates the critical boundary $K_{eff} = 1$. The region to the left (dark blue, $K_{eff} \ll 1$) represents the  safe parameter space where the near-threshold kinematics enforce $T_R \ll M_N$, resulting in a severe exponential suppression of the inverse-decay washout processes.}
\label{fig:washout_keff}
\end{figure}

In our numerical analysis we fix, for illustration, $M_N=5~\text{TeV}$ and use
the threshold relation $M_\chi\simeq 2M_N+\delta$ to obtain $T_R\sim 1~\text{TeV}$
for correlated values of $(Y_N,\delta)$. Figure~\ref{fig:TR_contours} shows the
parameter combinations that satisfy $T_R=1~\text{TeV}$, while
Fig.~\ref{fig:washout_keff} displays the corresponding region where the correct
baryon asymmetry is achieved once the resonant condition for CP enhancement is
imposed.

In summary, non-thermal leptogenesis from $\chi$ decay can overcome the thermal
catastrophe of the ISS $B\!-\!L$ model by enforcing $T_R<M_N$ (thereby suppressing
washout) while retaining a sufficiently large RHN yield and branching ratio. The
threshold ($\delta$) regime provides an efficient and technically natural way to
achieve this separation of scales, and it sets the stage for a successful
realization of baryogenesis in the inverse-seesaw $B\!-\!L$ model.
.

 While the parametric relation in Eq. (\ref{eq:etaB_substituted}) highlights the fundamental dilution-washout tension, it relies on standard thermal efficiency approximations. In a genuinely non-thermal setup, the heavy-neutrino abundance is not thermal, and the washout depends on the detailed time-evolution of the distributions. To establish the viability of this mechanism quantitatively, we must move beyond these estimates and treat the system dynamically.
\section{Dynamical Evolution and Boltzmann Equations}

In our previous parametric estimates, the survival of the baryon asymmetry relied on the assumption of an effective efficiency factor to model the washout suppression. To rigorously establish the viability of this non-thermal leptogenesis scenario and accurately track the energy transfer during the scalar decay, we now perform a full numerical integration of the coupled Boltzmann equations.

We track the proper time evolution of the comoving energy density of the decaying scalar field $\rho_\chi$, the radiation bath $\rho_R$, the number density of the heavy pseudo-Dirac neutrinos $n_N$, and the $B-L$ asymmetry $n_{B-L}$. The system is governed by:
\begin{align}
    \frac{d\rho_\chi}{dt} + 3H\rho_\chi &= -\Gamma_\chi \rho_\chi \\
    \frac{d\rho_R}{dt} + 4H\rho_R &= (1 - \text{Br}_N) \Gamma_\chi \rho_\chi + M_N \Gamma_N (n_N - n_N^{eq})\\
    \frac{dn_N}{dt} + 3Hn_N &= 2 \text{Br}_N \frac{\Gamma_\chi \rho_\chi}{M_\chi} - \Gamma_N (n_N - n_N^{eq})\\
    \frac{dn_{B-L}}{dt} + 3H n_{B-L} &= \epsilon_{CP} \Gamma_N (n_N - n_N^{eq}) - \Gamma_W n_{B-L}\
    \label{eq:boltzmann_NBL}
\end{align}
$\Gamma_\chi$ is the total decay width of the scalar field $\chi$, $\text{Br}_N$ is the branching ratio for the scalar decaying into a pair of heavy neutrinos ($\chi \rightarrow NN$). Consequently, $(1 - \text{Br}_N)$ is the fraction of the scalar's energy that decays directly into the standard radiation bath. $\Gamma_N$ is the total decay width of the heavy neutrinos. $n_N^{eq}$ is the equilibrium number density of the heavy neutrinos at a given temperature $T$. For non-relativistic neutrinos ($M_N \gg T$), this follows a Maxwell-Boltzmann distribution proportional to $T^{3/2} e^{-M_N/T}$. $\epsilon_{CP}$ is the resonant CP asymmetry (see Appendix \ref{sec:appendix_A} for the explicit Yukawa texture realization)

To close the system, the Hubble expansion rate $H$ is determined by the first Friedmann equation. Since the heavy neutrinos are strictly non-relativistic in our regime ($T \ll M_N$), their energy density can be approximated by their rest mass times their number density ($\rho_N \simeq M_N n_N$), leading to:
\begin{equation}
    H = \sqrt{\frac{8\pi}{3 M_{Pl}^2} (\rho_\chi + \rho_R + M_N n_N)}
\end{equation}
The temperature of the thermal bath, $T(t)$, is dynamically extracted from the radiation energy density at each integration step, $\rho_R(t) = (\pi^2/30) g_* T(t)^4$

The most critical aspect of the non-thermal evolution is the dynamical washout rate, $\Gamma_W$. Driven primarily by inverse decays ($L H \to N$), detailed balance in the thermal bath dictates that the macroscopic washout rate is proportional to the heavy neutrino decay width $\Gamma_N$, weighted by the thermodynamic penalty of extracting a heavy state from the bath \cite{Buchmuller:2004nz}:
\begin{equation}
\Gamma_W(t) = \frac{1}{2} \Gamma_N \frac{n_N^{eq}(T)}{n_\ell^{eq}(T)}
\end{equation}
where $n_\ell^{eq}(T) = (3/4)(\zeta(3)/\pi^2) g_\ell T^3$ is the equilibrium density of the massless Standard Model leptons, and $n_N^{eq}(T)$ is the equilibrium density of the heavy neutrinos, which follows a Maxwell-Boltzmann distribution $n_N^{eq} \propto T^{3/2} e^{-M_N/T}$.

By continuously tracking $T(t)$ through $\rho_R$, our numerical
integration captures the dynamical evolution of the washout rate
throughout all phases of the evolution. The washout term $\Gamma_W$
is evaluated at the instantaneous temperature $T(t)$ at every
timestep, with no approximation made at any stage. The detailed
behaviour of $\Gamma_W/H$ across these phases is discussed in the
following subsection.
\begin{figure}
    \centering
    \includegraphics[width=0.8\linewidth]{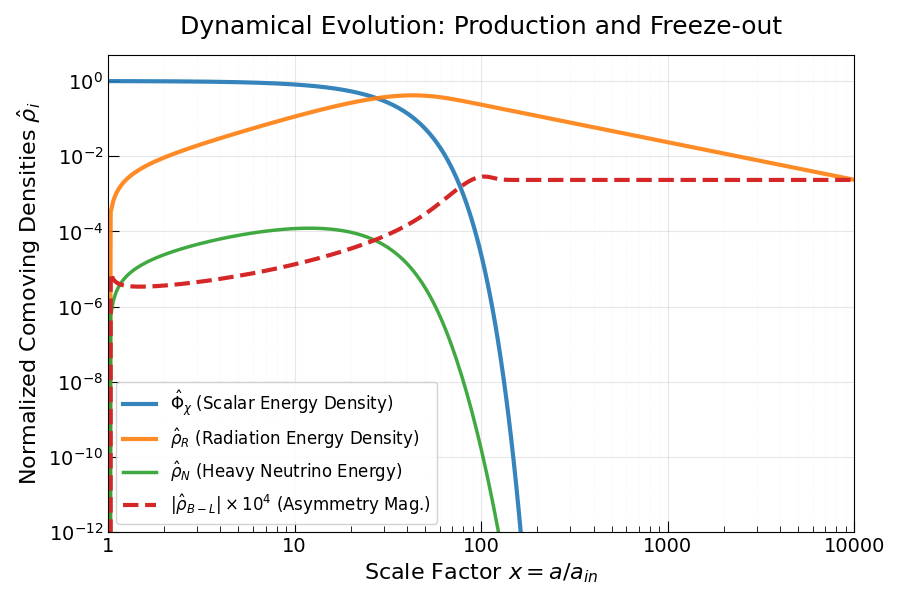}
    \caption{Dynamical evolution of comoving densities normalized by
    the initial comoving scalar energy density,
    $\Phi_{in} = \rho_{\chi}(x=1)$, tracking energy transfer as a
    function of the dimensionless scale factor $x = a/a_{in}$.
    The solid blue line shows the normalized comoving energy density
    of the decaying scalar field,
    $\hat{\Phi}_\chi(x) = (\rho_\chi x^3) / \Phi_{in}$.
    The solid orange line indicates the normalized comoving radiation
    energy density,
    $\hat{\rho}_R(x) = (\rho_R x^3) / \Phi_{in}$,
    illustrating the reheating process.
    The solid green line represents the normalized energy density
    stored in the non-relativistic heavy neutrinos,
    $\hat{\rho}_N(x) = (n_N \times M_N \times x^3) / \Phi_{in}$.
    Finally, the dashed red line depicts the absolute magnitude of
    the generated comoving $B-L$ energy density,
    $|\hat{\rho}_{B-L}(x)| = (|n_{B-L}| \times M_N \times x^3) /
    \Phi_{in}$, amplified by a factor of $10^4$ for visibility on
    this logarithmic scale. $a_{in}$ is the initial scale factor at
    the beginning of the decays. The parameters are fixed to the
    benchmark point: $M_N = 5$ TeV, $\delta = 10^{-6}$ GeV, and
    $y_\nu = 10^{-2}$. The $B-L$ asymmetry successfully freezes out
    and approaches a constant comoving value, confirming that the
    generated asymmetry is ultimately protected once
    $T \lesssim M_N$.}
    \label{fig_dynamical evolution}
\end{figure}

\subsection{Pre-reheating washout and the role of
             \boldmath{$T_{\rm max}$}}
\label{sec:Tmax}

A subtle but important point concerns the washout during the
pre-reheating phase, before radiation domination is established.
Although the reheating temperature satisfies $T_R \ll M_N$, the
bath temperature does not simply start at $T_R$. During the early
matter-dominated phase, the radiation bath is built up gradually by
the decay of $\chi$, and the temperature rises from its initial
value to a maximum $T_{\rm max}$ before falling back toward $T_R$
as the scalar energy is depleted. For our benchmark parameters
($M_N = 5$ TeV, $\delta = 10^{-6}$ GeV), we find numerically
$T_{\rm max} \approx 3.5\,M_N$, reached very early at
$x \approx 1.5$. This is a well-known feature of non-thermal
reheating scenarios \cite{Giudice:2000ex}: $T_{\rm max}$ can
significantly exceed $T_R$ even when $T_R \ll M_N$.

This raises the question of whether washout processes are active
during the pre-reheating phase $T_{\rm max} > T > T_R$. Our
numerical integration already addresses this question rigorously,
since the washout rate $\Gamma_W(t)$ is evaluated at the
instantaneous dynamical temperature $T(t)$, extracted from
$\rho_R(t)$ at every timestep, with no approximation made regarding
this phase.

Two
features are worth noting. First, the bath temperature peaks early
at $T_{\rm max} \approx 3.5\,M_N$ ($x \approx 1.5$), well above
$M_N$, so inverse decays are kinematically accessible during the
initial heating phase. Second, and more importantly,
$\Gamma_W/H$ itself does not peak at $T_{\rm max}$ but rather much
later, at $x \approx 4\times 10^3$ when $T \approx 0.2\,M_N$,
reaching $\Gamma_W/H \sim \mathcal{O}(10^7)$. This is because
$\Gamma_W/H \propto (M_N^2/T^2)\,e^{-M_N/T}/H$ grows as $H$
falls during matter domination, even while $T$ is already
decreasing, so the peak washout rate occurs deep in the cooling
phase rather than at $T_{\rm max}$.

Nevertheless, the generated $B-L$ asymmetry is not erased
throughout this entire period, for the following reason. The
evolution of $n_{B-L}$ is governed by the competition between two
terms in Eq.~(\ref{eq:boltzmann_NBL}): the CP-violating source
$\epsilon_{CP}\,\Gamma_N\,(n_N - n_N^{eq})$ and the washout term
$\Gamma_W\,n_{B-L}$. As long as $\chi$ is actively decaying and
injecting heavy neutrinos into the bath, $n_N \gg n_N^{eq}$, so
the source term dominates over the washout term.

Only after the scalar has largely decayed and $n_N$ approaches
$n_N^{eq}$ does the source switch off. At this point $T$ has
fallen sufficiently below $M_N$ that $\Gamma_W/H$ drops to zero
exponentially, and the final $B-L$ asymmetry freezes in. This
behaviour is confirmed in Fig.~\ref{fig_dynamical evolution}: the
comoving $B-L$ density reaches its asymptotic value and remains
constant once $T \ll M_N$, demonstrating that the late-time
exponential suppression of $\Gamma_W$ ultimately protects the
generated asymmetry.

\section{Resonant Leptogenesis in the Inverse Seesaw}

In the inverse seesaw model, the heavy neutrinos form quasi-degenerate
pseudo-Dirac pairs. The CP asymmetry generated from the mixing of two nearly
degenerate heavy neutrinos $N_1$ and $N_2$ is given by
\begin{equation}
\epsilon_{CP} \simeq
\frac{\mathrm{Im}\!\left[(h_{12})^2\right]}
{|h_{11}|^2 + |h_{22}|^2}
\frac{(M_1^2 - M_2^2)\, M_1 \Gamma_2}
{(M_1^2 - M_2^2)^2 + M_1^2 \Gamma_2^2},
\end{equation}
where $h = Y_\nu^\dagger Y_\nu$ and $\Gamma_2$ is the decay width of $N_2$.

Assuming maximal CP-violating phases, this expression simplifies to
\begin{equation}
\epsilon_{CP}^{\rm res} \simeq
\frac{\Delta M\, \Gamma}
{\Delta M^2 + \Gamma^2/4},
\end{equation}
where $\Delta M = M_2 - M_1$ and $\Gamma$ denotes the common decay width.
The CP asymmetry reaches its theoretical maximum,
\begin{equation}
\epsilon_{CP}^{\rm max} = \frac{1}{2},
\end{equation}
when the resonance condition
\begin{equation}
\Delta M \simeq \frac{\Gamma}{2}
\end{equation}
is satisfied. In appendix A, we shown how it is possible to get a texture for the Yukawa couplings that allowed to reach this maximal CP violation and still compatible with all light neutrinos datas. 

In the inverse seesaw, the mass splitting is controlled by the lepton-number
violating parameter $\mu$,
\begin{equation}
\Delta M \simeq \mu,
\end{equation}
while the decay width of a heavy neutrino is
\begin{equation}
\Gamma \simeq \frac{(y_\nu^\dagger y_\nu)}{8\pi} M_N.
\end{equation}
The resonance condition therefore implies
\begin{equation}
\mu \simeq \frac{y_\nu^2 M_N}{16\pi}.
\end{equation}

Since the Dirac Yukawa coupling is constrained by the light neutrino mass,
\begin{equation}
m_\nu \simeq \frac{(y_\nu v)^2}{M_N^2}\,\mu,
\end{equation}
one finds
\begin{equation}
y_\nu^2 \simeq \frac{m_\nu M_N^2}{v^2 \mu}.
\end{equation}
Substituting into the resonance condition yields the optimal mass splitting
\begin{equation}
\mu_{\rm res} \simeq
\sqrt{\frac{m_\nu M_N^3}{16\pi v^2}}.
\end{equation}

\begin{figure}
    \centering
    \includegraphics[width=0.75\linewidth]{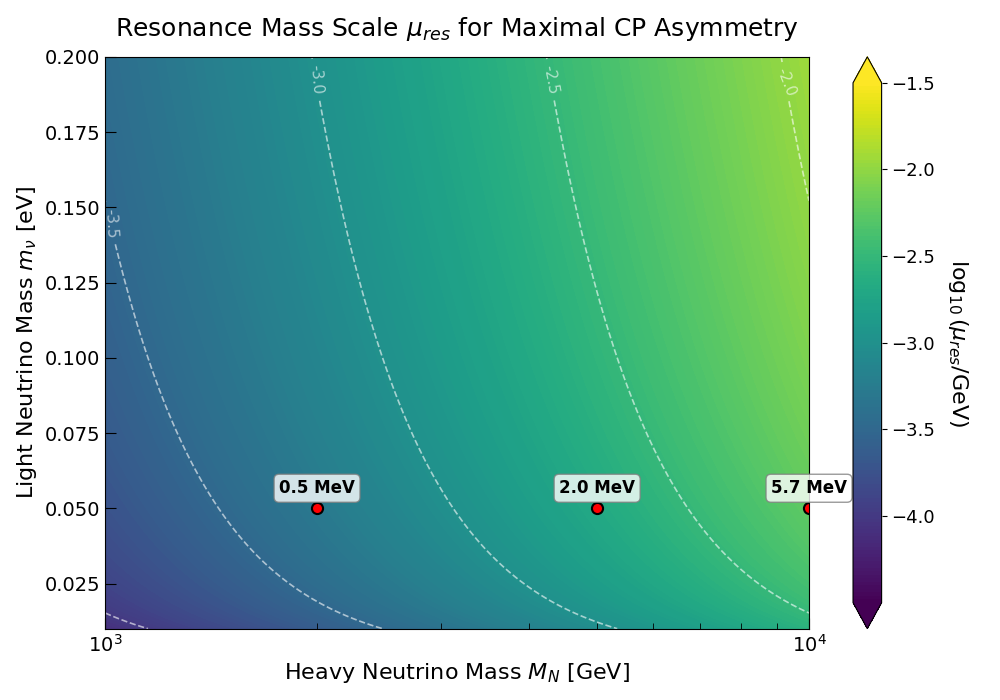}
    \caption{Contour plot illustrating the lepton-number-violating mass scale, $\mu_{res}$, required to satisfy the exact resonance condition ($\Delta M \simeq \Gamma_N / 2$) for maximal CP asymmetry. The required scale is shown as a function of the heavy pseudo-Dirac neutrino mass $M_N$ and the active light neutrino mass scale $m_\nu$. The color map displays $\log_{10}(\mu_{res} / \text{GeV})$, computed using the analytical tree-level Inverse Seesaw matching condition $\mu_{res} = \sqrt{m_\nu M_N^3 / (16 \pi v^2)}$, where $v = 174$ GeV is the SM Higgs vacuum expectation value. The red markers indicate specific benchmark scenarios for an atmospheric neutrino mass scale of $m_\nu = 0.05$ eV, demonstrating that for heavy neutrino masses in the multi-TeV range ($M_N = 2, 5$, and $10$ TeV), the necessary lepton-number-violating scale falls  in the $\mathcal{O}(\text{MeV})$ regime.}
    \label{fig:muresonance}
\end{figure}

\section{Numerical Results}

We now summarize the numerical results for the baryon asymmetry
$\eta_B$ obtained via non-thermal resonant leptogenesis in the
$U(1)_{B-L}$ inverse seesaw model, emphasizing the role of the threshold
mechanism in suppressing washout at the TeV scale.

\begin{figure}[t]
\centering
\includegraphics[width=0.8\linewidth]{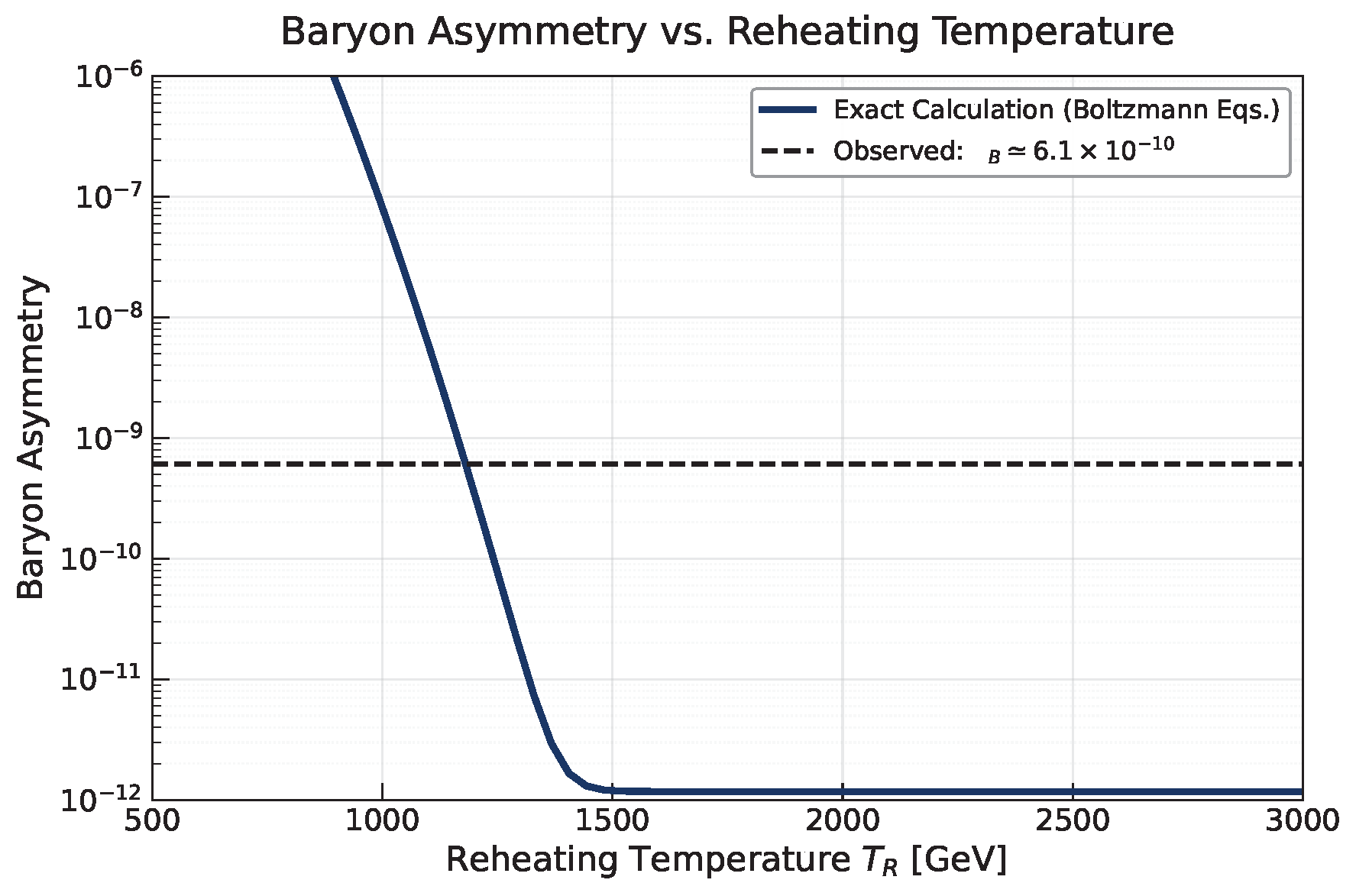}
\caption{\textbf{Dependence of the Baryon Asymmetry on the Reheating Temperature.} The final baryon-to-photon ratio $\eta_B$ (solid blue curve) is shown as a function of the reheating temperature $T_R$. The heavy neutrino mass is fixed to $M_N = 5$ TeV, with a Dirac Yukawa coupling of $y_\nu = 10^{-2}$ and a resonant CP asymmetry of $\epsilon_{CP} = 0.48$. The horizontal dashed line denotes the observed baryon asymmetry ($\eta_B \simeq 6.1 \times 10^{-10}$). As $T_R$ approaches $M_N$, the exponential suppression of the equilibrium density $n_N^{eq}(T)$ weakens, activating the inverse-decay washout processes ($\Gamma_W$) and driving the asymmetry to zero.}
\label{fig:eta_B_vs_TR}
\end{figure}

\begin{figure}
    \centering
    \includegraphics[width=0.75\linewidth]{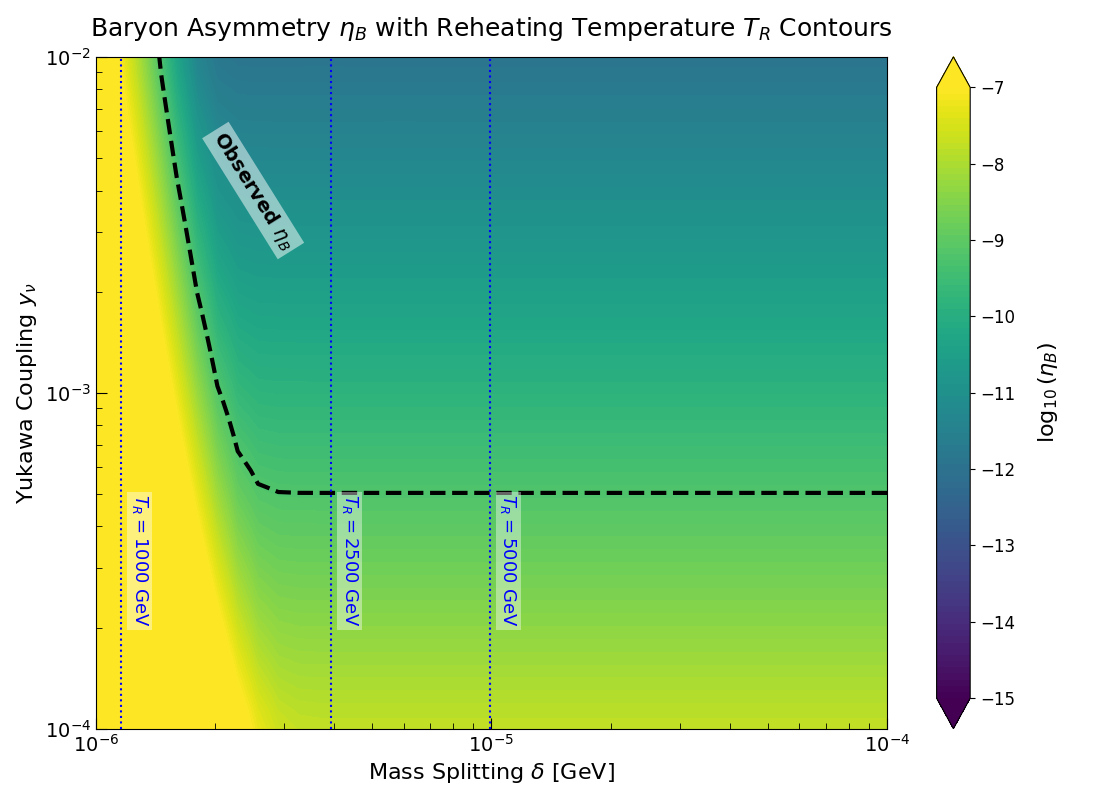}
   \caption{\textbf{Baryon Asymmetry from Full Dynamical Evolution.} Contour plot displaying the logarithmic magnitude of the final baryon-to-photon ratio, $\log_{10}(\eta_B)$, as a function of the scalar mass splitting $\delta$ and the Dirac Yukawa coupling $y_\nu$. The parameter space is  scanned over $\delta \in [10^{-6}, 10^{-4}]$ GeV and $y_\nu \in [10^{-4}, 10^{-2}]$. The remaining model parameters are fixed to our standard benchmark values: $M_N = 5$ TeV, scalar-neutrino coupling $Y_N = 1.0$, branching fraction $\text{Br}(\chi \to NN) = 0.5$, and the resonant CP asymmetry $\epsilon_{CP} = 0.48$. Every point is computed by extracting the asymptotic $B-L$ yield from the exact numerical integration of the fully coupled Boltzmann equations (Sec. IV). The thick black dashed contour marks the parameter space that successfully reproduces the observed baryon asymmetry of the Universe ($\eta_B \simeq 6.1 \times 10^{-10}$). The vertical blue dotted lines indicate contours of constant reheating temperature ($T_R = 1000$, $2500$, and $5000$ GeV), visually confirming that the target asymmetry is  achieved in the near-threshold regime ($\delta \sim \mathcal{O}(10^{-6})$ GeV) where $T_R$ is strictly bounded below the heavy neutrino mass scale, thereby dynamically evading thermal washout.}
\label{fig:eta_B_contours}
\end{figure}

\begin{itemize}
\item The reheating temperature $T_R$ is generated by the decay of the
$B\!-\!L$ scalar $\chi$ into heavy neutrinos. To avoid thermal washout, we impose
the non-thermal condition $T_R < M_N$.

\item Resonant enhancement of the CP asymmetry is achieved through the small
lepton-number violating parameter $\mu$:
\begin{itemize}
    \item For $M_N = 5$~TeV, the resonance condition is satisfied at
    $\mu \simeq 2$~MeV.
    \item Under these conditions, the CP asymmetry reaches
    $\epsilon_{CP} \simeq 0.48$.
\end{itemize}
\end{itemize}
The numerical results can be found in figures (\ref{fig:muresonance},\ref{fig:eta_B_vs_TR},\ref{fig:eta_B_contours}).



\subsection*{Phenomenological Constraints on the Benchmark Scenario}
\label{sec:constraints}
 
 Here, we  evaluate our successful leptogenesis benchmark point ($M_N = 5$ TeV, $y_\nu = 10^{-2}$) against the relevant experimental constraints:

\textbf{Heavy-Neutrino Mixing:} The active-heavy mixing parameter governs the non-unitarity of the PMNS matrix and direct collider production. 
For our benchmark, this yields $\theta \simeq y_\nu v / (\sqrt{2} M_N) \simeq 1.2 \times 10^{-4}$. The corresponding non-unitarity measure $\theta^2 \sim 10^{-8}$ is several orders of magnitude below current electroweak precision constraints, which require $\theta^2 \lesssim 10^{-3}$.

\textbf{Lepton Flavor Violation (LFV):} The most restrictive flavor constraint in the ISS is the radiative decay $\mu \to e \gamma$. The branching ratio depends on the used Casas-Ibarra matrix (as discussed in Appendix \ref{sec:appendix_A}). With an active-heavy mixing of $\theta \sim 10^{-4}$ and $M_N = 5$ TeV, the induced LFV rates are below the current MEG limit of $\text{Br}(\mu \to e \gamma) < 4.2 \times 10^{-13}$, though they may fall within the reach of the future MEG II upgrade.

\textbf{The $Z'$ Gauge Boson:} The $U(1)_{B-L}$ symmetry breaking generates a massive $Z'$ boson. Current ATLAS and CMS dilepton resonance searches constrain $M_{Z'} \gtrsim 4.5$ TeV for a gauge coupling $g_{B-L} \sim 0.1$. The $Z'$ mass is determined by the $B-L$ breaking scale $v_{B-L}$, which also sets the heavy neutrino mass. Choosing $v_{B-L}$ such that $M_{Z'} \sim 6$ TeV  evades LHC bounds without modifying the leptogenesis dynamics, as the $Z'$ decoupled from the thermal bath earlier in the cosmic history.

\textbf{Extended Higgs Sector:} The physical scalar $\chi$ mixes with the Standard Model Higgs boson $h$ with a mixing angle $\alpha$. To respect the LHC measurements of the 125 GeV Higgs signal strengths, $\sin \alpha \lesssim 0.1$ is needed. In our model, the near-threshold condition ($\delta \ll M_N$)  assumes a suppressed coupling between the scalar sector and the heavy neutrinos, making a  suppressed scalar mixing angle  consistent and natural.

In conclusion, in our non-thermal leptogenesis scenario,  the parameter space  is among the  allowed parameter spaces regions by current theoretical and experimental requirements.

\section{Conclusions}

We have shown that the gauged $U(1)_{B-L}$ inverse seesaw model provides a phenomenologically viable model for baryogenesis through non-thermal resonant leptogenesis. Although the large Yukawa couplings inherent to the inverse seesaw constrain standard thermal leptogenesis due to strong washout processes, this difficulty is  overcome by producing the heavy pseudo-Dirac neutrinos out of equilibrium via the decay of the $B\!-\!L$ Higgs scalar.

Two mechanisms are needed to get a successful baryogenesis in this scenario. First, setting the scalar mass close to the kinematic threshold, $M_\chi \simeq 2M_N + \delta$,  lowers the maximum temperature of the radiation bath to $T_R \ll M_N$ (while $T_{\rm max}$ can transiently exceed $M_N$, as discussed in Sec.~\ref{sec:Tmax}). The CP-violating source dominates over washout while $\chi$ is actively decaying, and once the source switches off the bath temperature has fallen sufficiently below $M_N$ that the washout rate becomes exponentially suppressed, protecting the frozen-in asymmetry. Second, the quasi-degenerate structure of the heavy neutrinos enables resonant CP violation. By employing a modified Casas-Ibarra parameterization, we confirmed that the CP asymmetry can reach macroscopic values ($\epsilon_{CP} \sim 0.45$) without suffering from  flavor cancellations when the resonance condition $\Delta M \simeq \Gamma_N/2$ is met.

Our exact numerical analysis demonstrates that for heavy neutrino masses $M_N \sim 5~\mathrm{TeV}$, reheating temperatures of order $T_R \sim 1~\mathrm{TeV}$, and a lepton-number-violating mass scale $\mu_S \sim \mathcal{O}(\mathrm{MeV})$, the observed baryon asymmetry of the Universe is successfully reproduced. Maintaining the narrow near-threshold condition ($\delta \sim 10^{-6}$ GeV) requires a parameter tuning of roughly one part in $10^5$ against one-loop radiative corrections. However, this degree of tuning is  quite common for TeV-scale phenomenological models and is  less severe than the electroweak hierarchy problem.

The scale of this scenario of non-thermal leptogenesis is   the TeV scale. We have verified that the required parameter space  evades current phenomenological constraints from lepton flavor violation (e.g., $\mu \to e \gamma$), LHC $Z'$ resonance searches, and active-heavy neutrino mixing bounds. This model provides a consistent connection between neutrino mass generation and the cosmological origin of matter, compatible with current data and open to future experimental scrutiny.

\begin{acknowledgments}
D.D. acknowledges financial support from SECIHTI, SNII, and U. of Gto. (M\'exico).
\end{acknowledgments}

\appendix
\appendix
\section{Concrete Yukawa Textures and Resonant CP Asymmetry}
\label{sec:appendix_A}

In pseudo-Dirac systems characteristic of the Inverse Seesaw (ISS) mechanism, the resonant CP asymmetry can suffer from severe cancellations. In this appendix, we explicitly demonstrate that our phenomenological benchmark of an unsuppressed resonant CP asymmetry ($\epsilon_{CP} \sim \mathcal{O}(0.1-0.5)$) is possible without fine-tuning the flavor structure, provided that the heavy neutrino mixing matrix contains $\mathcal{O}(1)$  phases.

The  active neutrino mass matrix in the ISS nodel is given by the exact tree-level formula:
$$m_\nu \simeq \frac{v^2}{2} Y_\nu M_N^{-1} \mu_S (M_N^T)^{-1} Y_\nu^T$$
where $v = 174$ GeV is the Standard Model Higgs vacuum expectation value, $M_N$ is the heavy Dirac mass matrix, and $\mu_S$ is the lepton-number-violating mass matrix. We adopt a modified Casas-Ibarra parameterization suitable for the ISS:
$$Y_\nu = \frac{\sqrt{2}}{v} U_{PMNS} \sqrt{m_\nu^{diag}} R \sqrt{\mu_S^{-1}} M_N$$
Here, $U_{PMNS}$ is the unitary neutrino mixing matrix, $m_\nu^{diag}$ contains the light active neutrino masses, and $R$ is an arbitrary complex orthogonal matrix ($R R^T = \mathbb{I}$) that encodes the high-scale CP-violating phases. 

The  CP asymmetry for the decay of the heavy states $\nu_{H_i}$ is due to the interference of tree-level and one-loop self-energy diagrams. In the resonant regime ($\Delta M \simeq \Gamma_N / 2$), the asymmetry is proportional to the interference term:
$$\epsilon_{CP} \propto \frac{\text{Im}[(Y_\nu Y_\nu^\dagger)_{12}^2]}{(Y_\nu Y_\nu^\dagger)_{11} (Y_\nu Y_\nu^\dagger)_{22}}$$

To avoid the catastrophic pseudo-Dirac cancellations, the matrix $R$ must be complex. We consider a minimal scenario with two heavy pseudo-Dirac pairs, defining $R$ with a complex angle $\theta = a + ib$:
$$R = \begin{pmatrix} \cos(a+ib) & \sin(a+ib) \\ -\sin(a+ib) & \cos(a+ib) \end{pmatrix}$$

For typical $\mathcal{O}(1)$ values of the parameters (e.g., $a \simeq \pi/4$ and $b \simeq 1.2$), the imaginary components of $R$ prevent the elements of $Y_\nu Y_\nu^\dagger$ from becoming strictly real or aligned. 

Inserting our benchmark values, $M_N = 5$ TeV and $\mu_S = 2$ MeV into the Casas-Ibarra parameterization yields elements of the Dirac Yukawa matrix of order $|(Y_\nu)_{ij}| \sim 10^{-2}$. 

Evaluating the exact CP asymmetry with this texture produces a non-vanishing interference term $\text{Im}[(Y_\nu Y_\nu^\dagger)_{12}^2] \neq 0$, yielding a macroscopic resonant asymmetry of:
$$\epsilon_{CP} \simeq 0.45$$

This explicit flavor realization confirms that the unsuppressed CP asymmetry utilized in our  Boltzmann integration is consistent with the stringent low-energy neutrino oscillation physics.

%

\end{document}